\documentclass[journal]{IEEEtran}
\usepackage{graphicx}
\begin{document}
\title{A DAQ Prototype for Front-end Waveform Digitization in Intensive Electromagnetic Field Circumstance}
%
%

\author{Zhou He, Lian Chen, Feng Li, Weigang Yin and Ge Jin
\thanks{ Manuscript received June 5,2018.
This work was supported by the National Natural Science Foundation of China under Grants 11461141010 and 11375179,
and in part by ``the Fundamental Research Funds for the Central Universities'' under grant No. WK2360000005
and the National key foundation for exploring scientific instrument of China under Grant No.2013YQ04086102.}%
\thanks{Zhou He, Lian Chan, Feng Li, Weigang Yin and Ge Jin are with State Key Laboratory of Particle Detection and Electronics, University of Science and Technology of China, Hefei, Anhui 230026, P.R. of China (phone: +86-551-63607152; e-mail: hezhou@mail.ustc.edu.cn, chenlian@ustc.edu.cn,   phonelee@ustc.edu.cn, yd1105@mail.ustc.edu.cn, goldjin@ustc.edu.cn).}%
\thanks{First author: Zhou He, Corresponding author: Lian Chen.}%
}

\maketitle
\thispagestyle{empty}

\begin{abstract}
A front-end waveform digitization data acquisition system prototype for pulsed magnetic field generator in inertial confinement fusion is described. The pulse magnetic field is created by discharging a high-voltage capacitor through a small wire-wound coil, and the Rogowski coil is used to measure the discharge current which can describe the corresponding magnetic field waveforms. The prototype is designed to measure the signal instead of the oscilloscope through a long-distance coaxial, which is greatly affected by electromagnetic interference caused by high-power loser. The outfield test result shows that the prototype can has a comparable performance as the oscilloscope for pulse magnetic field measurement.
\end{abstract}


\section{Introduction}

\IEEEPARstart{p}{lasma} confinement and the suppression of energy transport are fundamental to achieving the high-energy-density conditions necessary for fusion applications. Magnetizing the hot spot in an inertial confinement fusion (ICF) implosion can reduce conductive energy transport \cite{chang2011fusion}\cite{gotchev2009seeding}, thus increasing the ion temperature as well as the neutron yield. Due to the short duration of ICF experiment, we choose the pulsed magnetic field which is easy to generate compared with the steady-state magnetic field. The pulse magnetic field is created by discharging a high-voltage capacitor through a small wire-wound coil, and the Rogowski coil is used to measure the discharge current which can describe the corresponding magnetic field waveforms. Traditionally, the signal from the coil is transmitted to the oscilloscope through a long-distance coaxial cable because the intense electromagnetic radiation poses great threats to the nearby electronic systems.
Therefore, we designed a front-end DAQ prototype to replace the traditional measurement method. Accordingly, the Graphical User Interface (GUI) is written based on LabVIEW application platform, achieving the initialization and control of the DAQ prototype.

\begin{figure}[!t]
\centering
\includegraphics[width=3.5in]{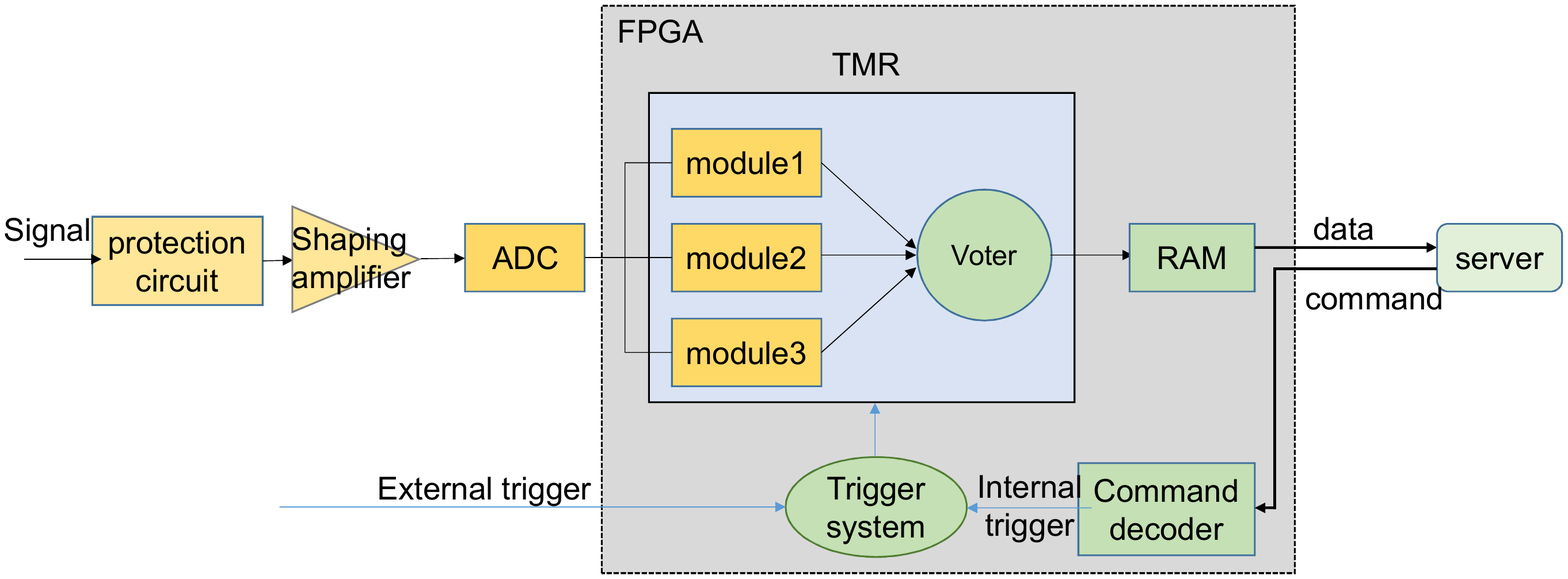}
\caption{Block diagram of the front-end signal digital acquisition system prototype}
\label{fig_1}
\end{figure}

\section{System Design}
\begin{figure}[!t]
\centering
\includegraphics[width=3.5in]{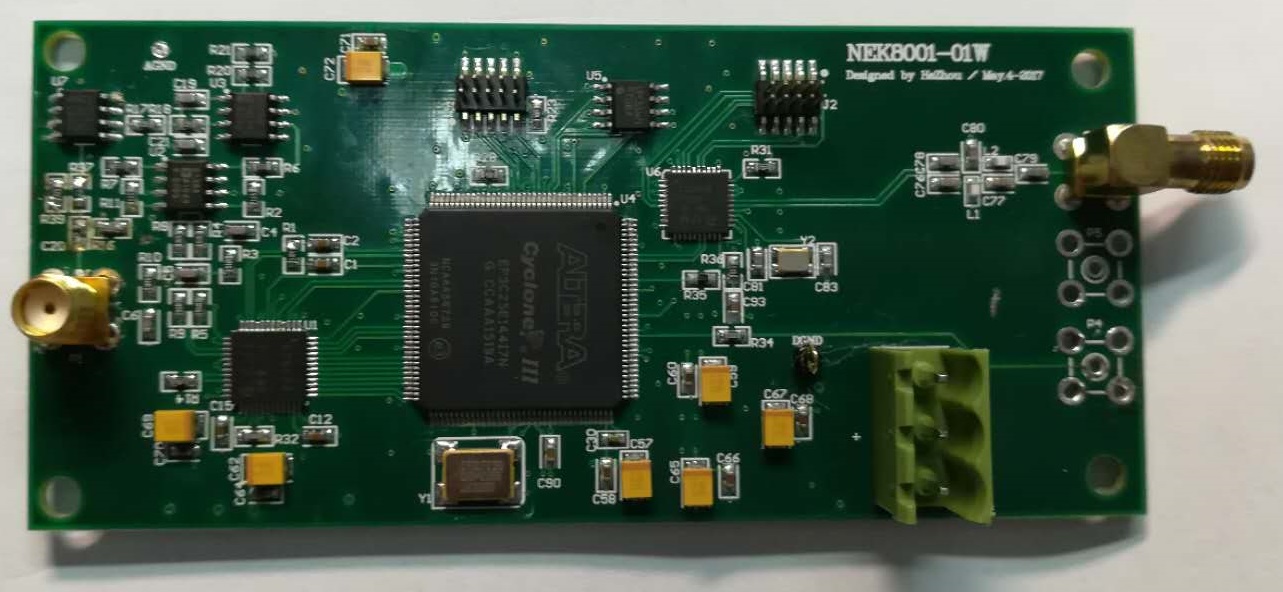}
\caption{the photo of the DAQ prototype}
\label{fig_2}
\end{figure}

Fig. \ref{fig_1} is a simplified block diagram of the DAQ for front-end waveform digitization and Fig. \ref{fig_2} is the photo of the DAQ prototype. The prototype size is 11cm $\times$ 5cm. The power consumption of the board is only 4.2W under $\pm$7V power supply.

\subsection{Hardware}

The core of the prototype is based on a cyclone-\uppercase\expandafter{\romannumeral3} FPGA, which is configured by a Serial Peripheral Interface (SPI) flash. Since the coil bandwidth is $\sim$3MHz \cite{fiksel2015note}, according to Nyquist sampling theorem, we choose a high speed FADC which has a sample rate of up to 210MSPS. The ESD protection circuit based on TVS transient diode is designed to reduce transient strong electromagnetic interference in ICF experiment. Since signal generated by the Rogowski coil is large, a type attenuation network is used in order to ensure the signal amplitude within the dynamic range of ADC.

\subsection{Signal flow in FPGA}

The block diagram of the signal flow in the FPGA is shown in the middle of Fig. \ref{fig_1}. The user-defined commands are sent from GUI and derived from the command decoder. At initialization, all the FIFOs and status registers are reset. Then a command for trigger waiting arrives. The system provides two trigger modes, external trigger mode for synchronization with the ICF experiment and internal trigger mode when the signal amplitude is known. The triple modular redundancy (TMR), which can reduce the impact of signal-event upset in strong magnetic field environment, is used in FPGA logic design. Since the electromagnetic interference will greatly affect the signal transmission, the wave signal is stored in the RAM and transmitted to the server after the implosion process.

\begin{figure}[!t]
\centering
\includegraphics[width=3.5in]{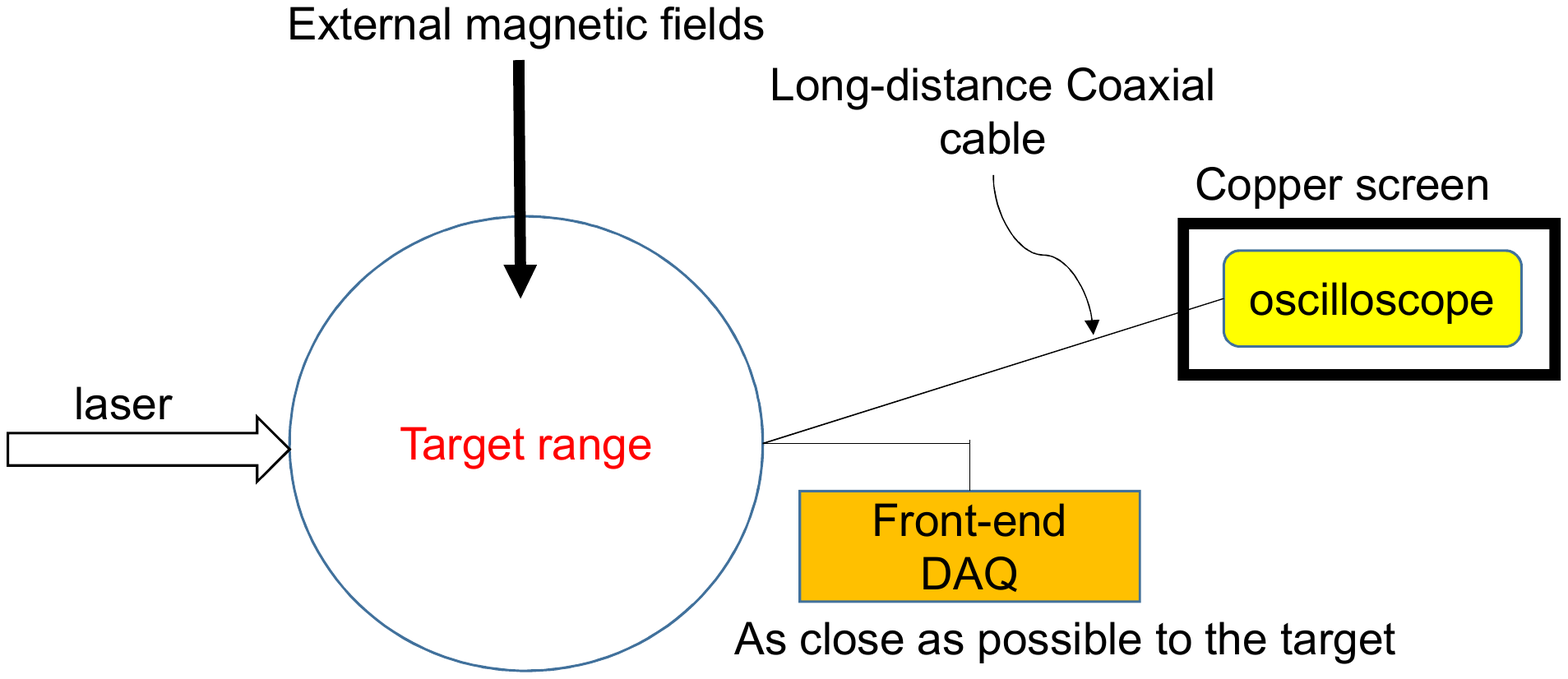}
\caption{the test structure}
\label{fig_3}
\end{figure}

\section{Experimental Result}
In order to test the performance of the prototype, we conducted the test on the experimental site of strong electromagnetic radiation. This experiment adds magnetic field to laser plasma, forming laser magnetized plasma.
The prototype is placed at the front end, as close as possible to the target chamber. Since the commercial oscilloscope does not have good electromagnetic shielding protection, it cannot be placed in the strong magnetic field directly. As a comparison, the signal is divided into two channels. The other channel is transmitted to the oscilloscope through 20m coaxial cable and placed in a copper shielding chamber. The test structure is shown in Fig. \ref{fig_3}.

\begin{figure}[!t]
\centering
\includegraphics[width=3.5in]{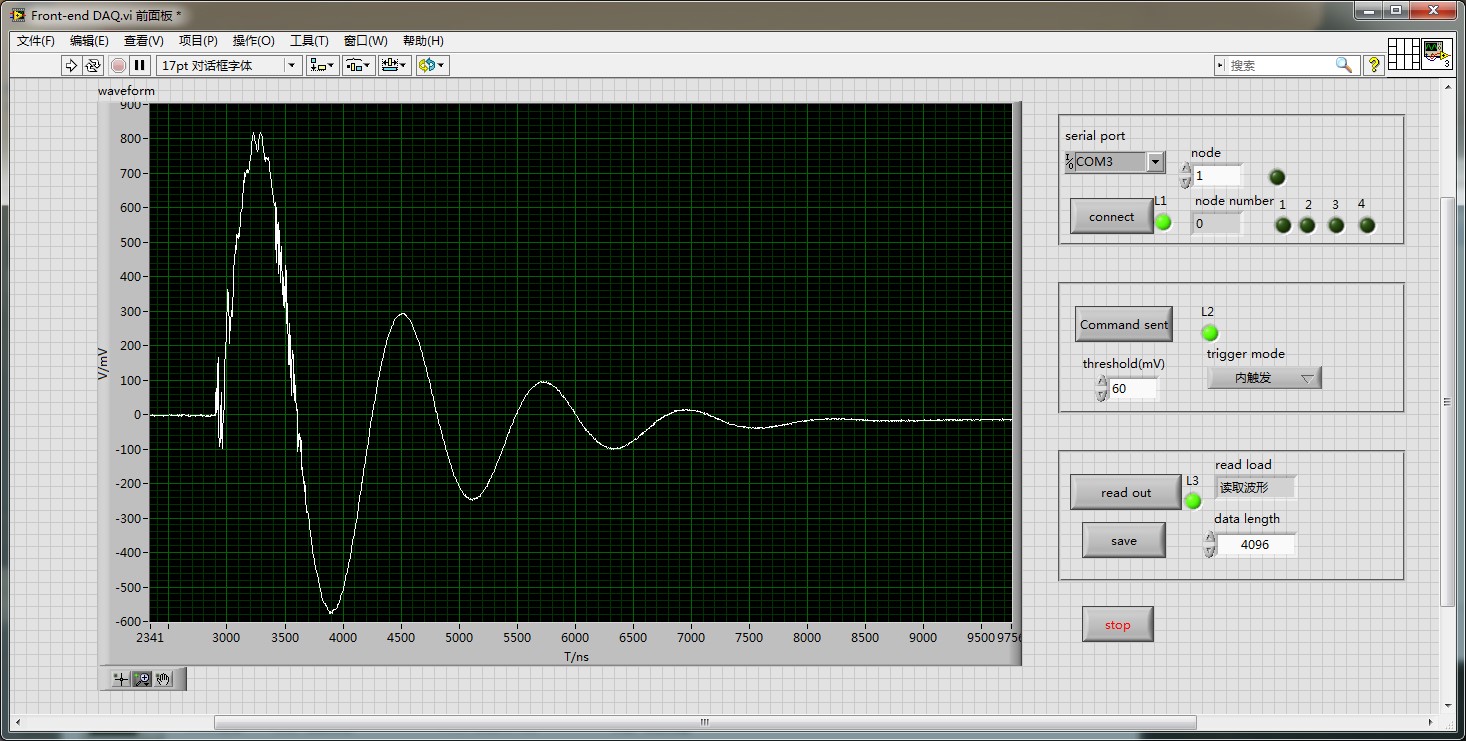}
\caption{the waveform from front-end DAQ}
\label{fig_4}
\end{figure}

\begin{figure}[!t]
\centering
\includegraphics[width=3.5in]{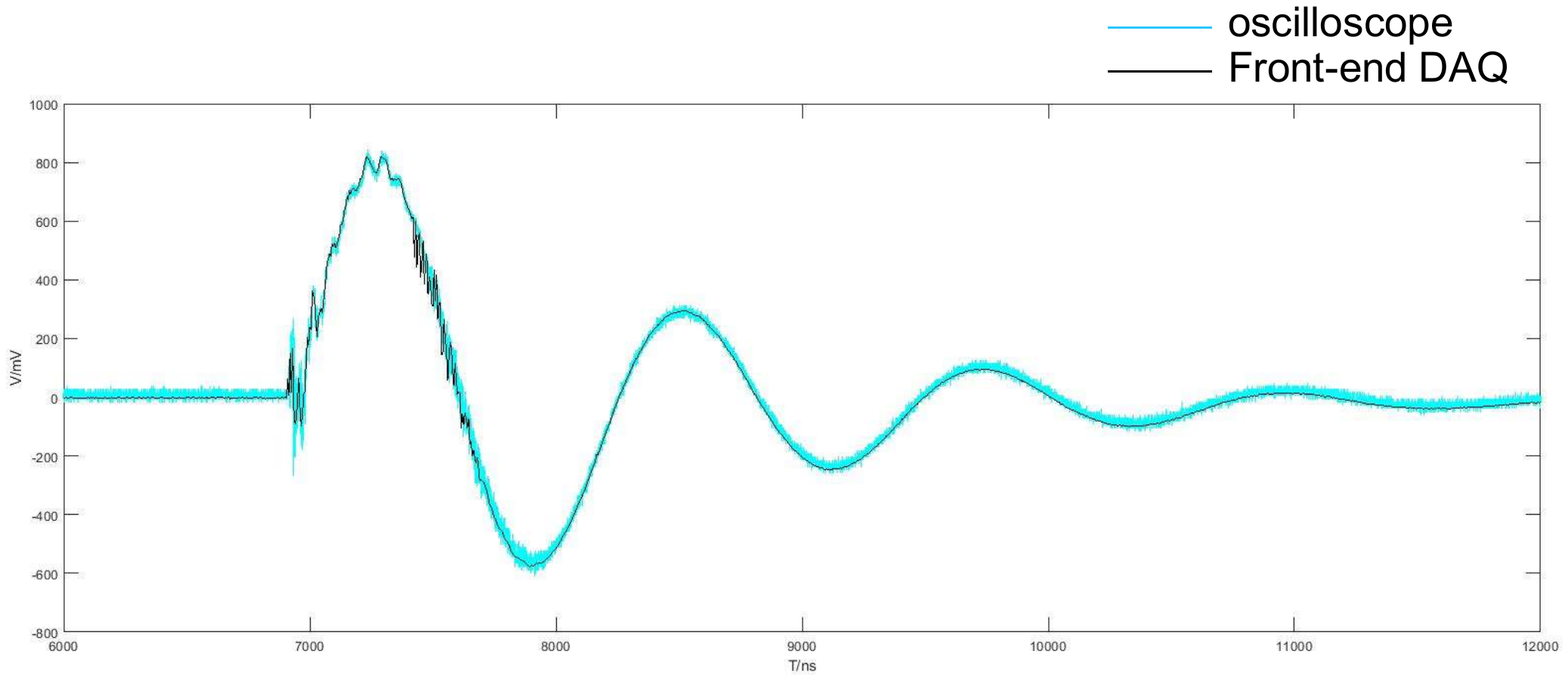}
\caption{the waveform compared between oscilloscope and front-end DAQ}
\label{fig_5}
\end{figure}

Fig. \ref{fig_4} shows the waveform measured by the prototype. After setting the trigger mode and waveform length, we issue the command through the upper computer. The comparison between the waveform measured by prototype and oscilloscope is shown in Fig. \ref{fig_5}. As can be seen from the figure, the prototype can still work normally under the strong magnetic field environment. By comparing the rising time and waveform amplitude, the waveform tested by prototype is consistent with the signal measured by oscilloscope. And that, the ADC sampling accuracy is 12 bits. Therefore, the prototype has better sampling accuracy and lower noise Compared with oscilloscope.

\section{Conclusion}

In this paper, we design a DAQ prototype for front-end waveform digitization in intensive electromagnetic field circumstance. It¡¯s proved through the field test that the prototype can replace the oscilloscope to read out the waveform in the intensive magnetic field environment and has an advantage in high sampling accuracy and low noise.

\end{document}